\documentclass[aps,amsfonts,amsmath,prd,preprint,nofootinbib]{revtex4}
\pdfoutput=1
\newcommand{\beq}{\begin{equation}}
\newcommand{\eeq}{\end{equation}}
\usepackage{graphicx}
\usepackage{natbib}
\usepackage[utf8]{inputenc}
\begin{document}

\title{A possible mass distribution of primordial black holes implied by LIGO-Virgo}
\author{Heling Deng}
\email{heling.deng@asu.edu}
\affiliation{Physics Department, Arizona State University, Tempe, AZ 85287, USA}

\begin{abstract}

The LIGO-Virgo Collaboration has so far detected around 90 black holes, some of which have masses larger than what were expected from the
collapse of stars. The mass distribution of LIGO-Virgo black holes
appears to have a peak at $\sim30M_{\odot}$ and two tails on the
ends. By assuming that they all have a primordial origin, we analyze
the GWTC-1 (O1\&O2) and GWTC-2 (O3a) datasets by performing maximum
likelihood estimation on a broken power law mass function $f(m)$,
with the result $f\propto m^{1.2}$ for $m<35M_{\odot}$ and $f\propto m^{-4}$
for $m>35M_{\odot}$. This appears to behave better than the
popular log-normal mass function. Surprisingly, such a simple and unique distribution can
be realized in our previously proposed mechanism of PBH formation,
where the black holes are formed by vacuum bubbles that
nucleate during inflation via quantum tunneling. Moreover, this mass distribution can also provide an explanation to supermassive black holes formed
at high redshifts.

\end{abstract}

\maketitle

\section{Introduction}

Over the past few years, the LIGO-Virgo Collaboration has detected gravitational
waves emitted by about 50 inspiraling and merging black hole binaries \cite{LIGOScientific:2018mvr,Abbott:2020niy},
and many more events are anticipated in the near future. It is intriguing
that most of the black holes have masses $\sim 30M_{\odot}$, which certainly
provides indicative information of the mass distribution of black
holes in the universe.

The origin of these black holes is so far unknown. While the possibility
that they are ordinary astrophysical black holes from stellar collapse (possibly from different
channels) are under active investigations \cite{Wong:2020ise,Zevin:2020gbd,Rodriguez:2021nwd,Fishbach:2021yvy}, a fascinating
speculation is that LIGO-Virgo has detected primordial black holes
(PBHs) \cite{Bird:2016dcv,Clesse:2016vqa,Sasaki:2016jop}. PBHs are hypothetical black holes formed by deviations from
density homogeneity in the early universe, before any large scale
structures and galaxies. Their masses can range from the Planck mass ($\sim 10^{-5}$ g)
to many orders of magnitude larger than the solar mass ($M_\odot \sim 10^{33}$ g). After the
release of LIGO-Virgo results, much effort has been dedicated to constrain
PBHs and the role they play in dark matter \cite{Sasaki:2016jop,Raidal:2017mfl,Ali-Haimoud:2017rtz,Vaskonen:2019jpv,Garriga:2019vqu}. It is now generally recognized
that they are unlikely to constitute all dark matter due to the small
merger rate inferred by LIGO-Virgo. 

It is also of great interest to constrain particular PBH mass distributions
with the LIGO-Virgo results \cite{Raidal:2017mfl,Raidal:2018bbj,Chen:2018czv,Gow:2019pok,Wu:2020drm}. Among them the log-normal mass function
\cite{Dolgov:1992pu} is the most popular because it is a good approximation for
various PBH mechanisms \cite{Clesse:2015wea,Blinnikov:2016bxu,Kannike:2017bxn}. However, if all LIGO-Virgo black holes
were PBHs, the peak of the log-normal function should be at $\sim20M_{\odot}$,
which is incompatible with what was observed. The possibility of
LIGO-Virgo black holes being a mixture of two populations of black
holes was recently considered in ref. \cite{Hutsi:2020sol}, which concludes that
astrophysical black holes that dominate the mass range $m\lesssim15M_{\odot}$ together with PBHs given by a log-normal (or critical collapse) mass function are much more probable than PBHs only.

In this paper, inspired by ref. \cite{Hutsi:2020sol}, we perform a maximum
likelihood estimation to fit the LIGO-Virgo datasets with a broken
power law only. More sophisticated techniques (e.g., Bayesian analysis
\cite{Hall:2020daa,Wong:2020yig}) are available, but we believe likelihood analysis should suffice
for our purposes, given the limited number of detected events and
uncertainties in PBH formation. We shall
also neglect some details in analyzing the formation of PBHs and binaries,
as long as they are not expected to affect the results by orders of
magnitude. For instance, following ref. \cite{Hutsi:2020sol}, we shall not consider
the impact from black hole's spin and mass accretion.

Apart from the single peak at $\sim30M_{\odot}$ in the distribution
of LIGO-Virgo black holes, our investigation of the broken power law
mass spectrum is mainly motivated by a PBH mechanism we proposed in
recent years. In a series of works \cite{Garriga:2015fdk,Deng:2017uwc,Deng:2020mds}, we studied PBHs formed by vacuum
bubbles that possibly nucleate during inflation. Depending on its
size after inflation ends, a bubble will turn into a black hole in
the either subcritical or supercritical regime, and the mass distributions of black holes
in these two regimes could obey different power laws. We would like
to know if the best-fit parameters from likelihood analysis of LIGO-Virgo data is compatible with this mechanism. Surprisingly, the broken power law mass function implied by the LIGO-Virgo black holes can indeed be realized in our model.

The rest of the paper is organized as follows. In section \ref{PBHs and LIGO-Virgo events}, we will first discuss the merger rate of PBHs and the detection probability of LIGO-Virgo, and then apply maximum likelihood estimation to find the best-fit parameters for the broken power law mass function. In section \ref{PBHs from primordial bubbles}, we will study the mechanism of PBH formation from primordial bubbles and how could produce the LIGO-Virgo black holes. Conclusions will be summarized and discussed in section \ref{Conclusions}. We set $c=\hbar=G=1$ throughout the paper.

\section{\label{PBHs and LIGO-Virgo events} PBHs and LIGO-Virgo events}

It will be assumed that all black holes detected by LIGO-Virgo are primordial.
They are formed during the radiation era and behave like dark matter,
with their number and mass density diluted by Hubble expansion. Two
neighboring black holes may collide as their gravitational attraction
defeats the Hubble stretching before the radiation-dust equality.
Disturbance from the surrounding environment, a typical example being
a third nearby black hole exerting a tidal torque, may impede the
head-on collision, leading to the formation of an inspiraling binary.
If the coalescence time of the binary is comparable to the age of
the universe, gravitational waves from the merger could possibly be
detected when they reach the earth. Roughly speaking, whether a merger
event can be heard depends on the detector's sensitivity, the masses
of the two black holes (source masses), the time when the merger took
place (source redshift), and the sky location and orientation of the
binary system. In order to estimate how often a merger event can be
recorded, we also need the merger rate of the binaries, which is determined by the underlying mechanism of PBH formation.

\subsection{PBH mass function and merger rate}

We characterize the mass distribution of PBHs by mass function $\psi(m)$,
defined by
\begin{equation}
\psi(m)=\frac{m}{\rho_{CDM}}\frac{\text{d}n}{\text{d}m}.\label{eq:psi def}
\end{equation}
Here $m$ is the black hole mass, $\text{d}n$ is the number density
of black holes within the mass range $(m,m+\text{d}m)$, and $\rho_{CDM}$
is the energy density of cold dark matter. Since black holes and dark
matter are diluted by cosmic expansion in the same way, $\psi(m)$
is a constant over time. Integrating $\psi(m)$ gives the total fraction
of dark matter in PBHs:
\begin{equation}
f_{PBH}\equiv\frac{\rho_{PBH}}{\rho_{CDM}}=\int\psi(m)\text{d}m,
\end{equation}
where $\rho_{PBH}=\int m\text{d}n$ is the energy density of PBHs.
If all dark matter is made of PBHs, we have $f_{PBH}=1$.

Another function often used to describe the PBH mass distribution
is
\begin{equation}
f(m)\equiv m\psi(m),\label{eq:f def}
\end{equation}
which can be interpreted as the fraction of dark matter in PBHs at
$m$ within the mass range $\Delta m\sim m$. This function is particularly
useful if the spectrum is relatively broad.

The PBH merger rate in the early universe is well studied in the literature \cite{Sasaki:2016jop,Raidal:2017mfl,Ali-Haimoud:2017rtz,Vaskonen:2019jpv,Garriga:2019vqu}.
In this paper we take the formula of differential merger rate from
ref. \cite{Raidal:2018bbj}: 
\begin{equation}
\text{d}R\approx\frac{1.6\times10^{6}}{\text{Gpc}^{3}\text{yr}}f_{PBH}^{-21/37}\left(\frac{t}{t_{0}}\right)^{-34/37}\left(\frac{M}{M_{\odot}}\right)^{-32/37}\eta^{-34/37}S(\psi,f_{PBH},M)\psi(m_{1})\psi(m_{2})\text{d}m_{1}\text{d}m_{2}.\label{eq:R}
\end{equation}
Integrating $\text{d}R$ over all masses gives the number of merger
per $\text{Gpc}^{3}$ per year. Here $m_{1}$ and $m_{2}$ are the
two masses in the binary, $M\equiv m_{1}+m_{2}$ and $\eta\equiv m_{1}m_{2}/M^{2}$;
$t$ is the time when the merger occurs, and $t_{0}$ is the present
time; $S$ is a suppression factor accounting for the effects from
other matter components, including nearby black holes, and $S=\mathcal{O}(1)$ for a not particularly wide spectrum. In our calculations
we shall approximate $S$ by a simplified expression given in ref.
\cite{Hutsi:2020sol}, but setting $S=1$ would not lead to much difference. 

To roughly estimate the largest $f_{PBH}$ we can have from the LIGO-Virgo
results, we can assume a monochromatic mass function with $\psi(m)=f_{PBH}\delta(m-30M_{\odot})$,
i.e., all PBHs have mass $30M_{\odot}$. Setting $t=t_{0}$ and $S=1$,
we have
\begin{equation}
\int\text{d}R\sim10^{5}f_{PBH}^{53/37}\text{Gpc}^{-3}\text{yr}^{-1}.
\end{equation}
The merger rate implied by LIGO-Virgo is $\mathcal{O}(10-100)\text{Gpc}^{-3}\text{yr}^{-1}$,
which gives $f_{PBH}\sim10^{-3}$. Therefore, if all PBHs are of masses
around $30M_{\odot}$, they can at most contribute to $0.1\%$ of
the dark matter.

\begin{figure}
\includegraphics[scale=0.3]{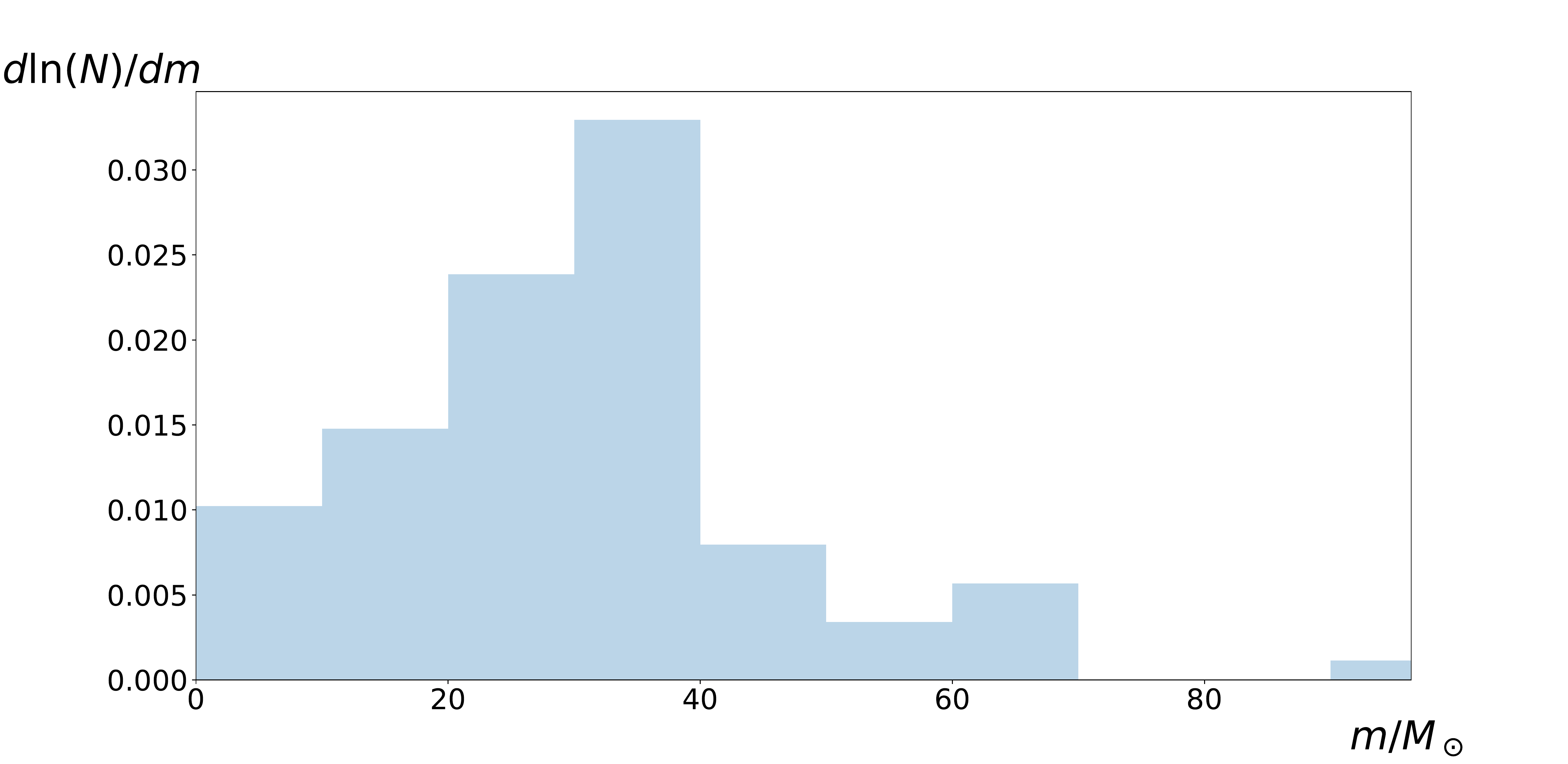}

\caption{\label{fig:dNdm}Numbers of detected LIGO-Virgo black holes in different mass
ranges, rescaled such that the areas add up to 1. There are 88 black
holes in total. The black hole masses are simply taken
to be the median values reported in refs. \cite{LIGOScientific:2018mvr,Abbott:2020niy}. The distribution certainly depends on the resolution of mass and the choice here is somehow arbitrary. Considering that the measurement uncertainties are $\mathcal{O}(10)M_\odot$ and that a smaller resolution would give a less clear pattern due to the insufficient number of detected events, we believe $\Delta m = 10 M_\odot$ is a plausible choice.}
\end{figure}

The distribution of LIGO-Virgo black holes is clearly not monochromatic.
In this paper we shall consider the 10 black hole merger events from
the GWTC-1 catalog \cite{LIGOScientific:2018mvr} and 34 events from GWTC-2 (discarding GW190719
and GW190909) \cite{Abbott:2020niy}, so there are 88 black holes in our dataset. In fig.
\ref{fig:dNdm}, we depict the black hole numbers in different mass
ranges. The distribution appears to have a peak at $m_{*}\sim30M_{\odot}$
and two tails on the ends, with more black holes having masses less
than $m_{*}$ than those on the other end. The simplest mass function
that can be conceived is a broken power law,
\begin{equation}
\psi(m)=\frac{f_{PBH}}{m_{*}}\left(\alpha_{1}^{-1}-\alpha_{2}^{-1}\right)^{-1}\begin{cases}
(m/m_{*})^{\alpha_{1}-1}, & m<m_{*}\\
(m/m_{*})^{\alpha_{2}-1}, & m>m_{*}
\end{cases},\label{eq:psi}
\end{equation}
which satisfies $\int\psi\text{d}m=f_{PBH}$. We further assume $\alpha_{1}>0$
and $\alpha_{2}<0$ such that $f(m)=m\psi(m)$ has a peak at $m_{*}$,
which means most contribution to the PBH energy density comes from
black holes with masses around $m_{*}.$ In the next subsection, we
will analyze the LIGO-Virgo results and find the best-fit parameters
for (\ref{eq:psi}). 

\subsection{Maximum likelihood estimation}

Given the differential merger rate found by eq. (\ref{eq:R}), the
expected number of merger signals reaching the earth per unit time
within the ranges $(m_{1},m_{1}+\text{d}m_{1})$, $(m_{2},m_{2}+\text{d}m_{2})$,
and $(z,z+\text{d}z)$ can be evaluated as \cite{Dominik:2014yma}
\begin{equation}
\text{d}N(m_{1},m_{2},z)=\frac{1}{1+z}\frac{\text{d}V_{c}}{\text{d}z}\frac{\text{d}R}{\text{d}m_{1}\text{d}m_{2}}\text{d}m_{1}\text{d}m_{2}\text{d}z,
\end{equation}
where $(1+z)^{-1}$ accounts for the time difference between the source
frame and the detector frame, and $V_{c}$ is the comoving Hubble
volume. For later convenience, we define
\begin{equation}
\Lambda(m_1,m_2,z)\equiv \frac{1}{1+z}\frac{\text{d}V_{c}}{\text{d}z}\frac{\text{d}R}{\text{d}m_{1}\text{d}m_{2}}
\end{equation}

In addition, we ought to take into account the fact that not all merger
events can be observed by LIGO-Virgo. The detection probability of
an event depends on the sensitivity of the instruments, the waveform
of the signal as well as the extrinsic parameters of the binary system,
i.e., its sky location and orientation. Integrating out the extrinsic
parameters for a given detector gives its detection probability $p_{det}(r),$
with $r\equiv\rho_{c}/\rho(m_{1},m_{2},z)$. Here $\rho_{c}$ is the
threshold signal-to-noise ratio above which a signal can be detected,
usually taken as $\rho_{c}=8$; and $\rho$ is the signal-to-noise
ratio for a merger located directly above the detector. $\rho$ is
defined by 
\begin{equation}
\rho(m_{1},m_{2},z)=2\sqrt{\int\frac{|\tilde{h}(f)|^{2}}{S_{n}(f)}\text{d}f},
\end{equation}
where $\tilde{h}(f)$ is the Fourier transform of the signal, and
$S_{n}(f)$ is the power spectrum of the detector's strain noise.
In this work, $\tilde{h}(f)$ and $S_{n}(f)$ are taken from refs.
\cite{Ajith:2007kx,Hutsi:2020sol} and refs. \cite{LIGOScientific:2018mvr,Abbott:2020niy} respectively. Then the detection probability of a particular event
 can be found by the parameter fit of $p_{det}(r)$ given in the appendix
of ref. \cite{Dominik:2014yma}. Since $S_{n}(f)$ is unique to a certain run, $p_{det}(r)$
takes different values in, e.g., O1\&O2 and O3a of LIGO-Virgo.

Now the total expected number of detection during a certain run can be found by
\begin{equation}
N_{e}=T_{o}\int p_{det}(m_{1},m_{2},z)\text{d}N,\label{eq:N}
\end{equation}
where $T_{o}$ is the observation time interval (166.6
days for O1\&O2 and 183.3 days for O3a). Therefore, once we are given the PBH mass function and the detector's
sensitivities ($S_{n}(f)$), we are able to obtain from eq. (\ref{eq:N})
the expected number of observable events. 

We assume that (observable) mergers occur randomly with a constant
rate, so they reach the detectors following a Poisson process. Let
$N_{o}$ be the actual number of observed events. Then the probability
of this happening during the observation interval is
\begin{equation}
p_{Poisson}\propto N_{e}^{N_{o}}e^{-N_{e}}.
\end{equation}
The likelihood of a single detected event with source masses $m_{1,o}$,
$m_{2,o}$ and redshift $z_{o}$ is
\begin{equation}
p_{o}\propto\frac{p_{det}(m_{1,o},m_{2,o},z_{o})\Lambda(m_{1,o},m_{2,o},z_{o})}{\int p_{det}(m_{1},m_{2},z)\text{d}N}.\label{eq:LD of single event}
\end{equation}
Here we have made an assumption that the source masses and redshift
can be perfectly determined by detection. In practice, we take $m_{1,o}$,
$m_{2,o}$ and $z_{o}$ to be the median values of the posterior samples
reported by LIGO-Virgo \cite{LIGOScientific:2018mvr,Abbott:2020niy}. In reality, their values are given
with probability distributions resulting from measurement uncertainties.\footnote{In order to account for the measurement uncertainties, the standard
approach to calculate $p_{o}$ is to replace the numerator in eq.
(\ref{eq:LD of single event}) by $\left\langle \Lambda(m_{1,o},m_{2,o},z_{o})/\pi(m_{1,o},m_{2,o},z_{o})\right\rangle $,
where $m_{1,o}$, $m_{2,o}$ and $z_{o}$ are the posterior samples,
$\pi$ is the corresponding source prior, and the brackets denote
an average over all samples of that event \cite{Mandel:2018mve,Hall:2020daa}.} Note that the denominator
in eq. (\ref{eq:LD of single event}) is $\propto N_{e}$. Then the
likelihood of all black hole merger events detected by LIGO-Virgo
is
\begin{align}
\mathcal{L} & =p_{Poisson}\prod_{i=1}^{N_{o}}p_{o}^{i}\propto e^{-N_{e}}\prod_{i=1}^{N_{o}}p_{det}(m_{1,o}^{i},m_{2,o}^{i},z_{o}^{i})\Lambda(m^i_{1,o},m^i_{2,o},z^i_{o}),
\end{align}
where each event is denoted by a superscript $i$. By maximizing the log-likelihood $\ln \mathcal{L}$ with the GWTC-1 and GWTC-2 datasets,
we will be able to attain the constraints on a PBH mass function.

\begin{figure}
\includegraphics[scale=0.22]{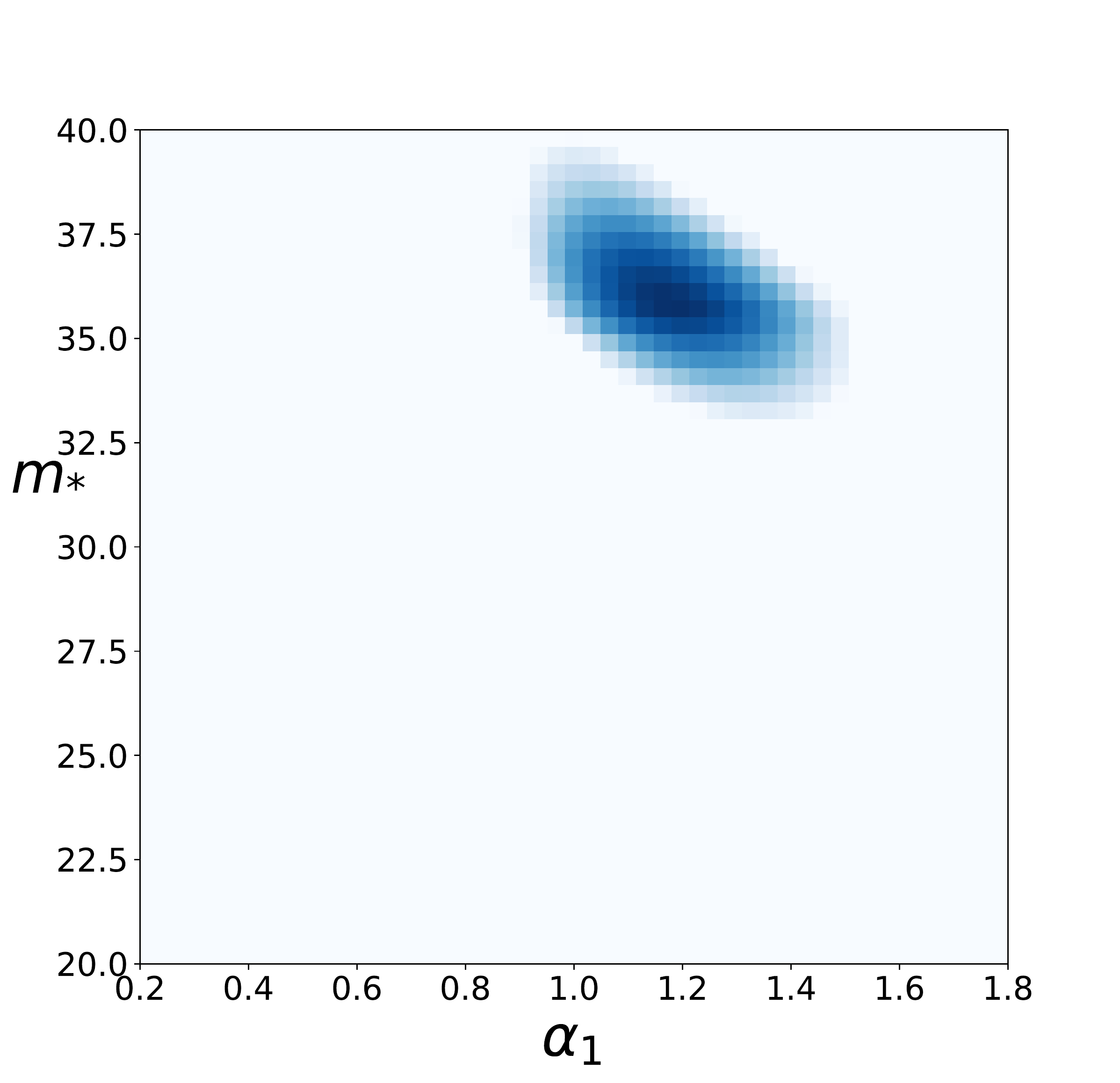}\includegraphics[scale=0.22]{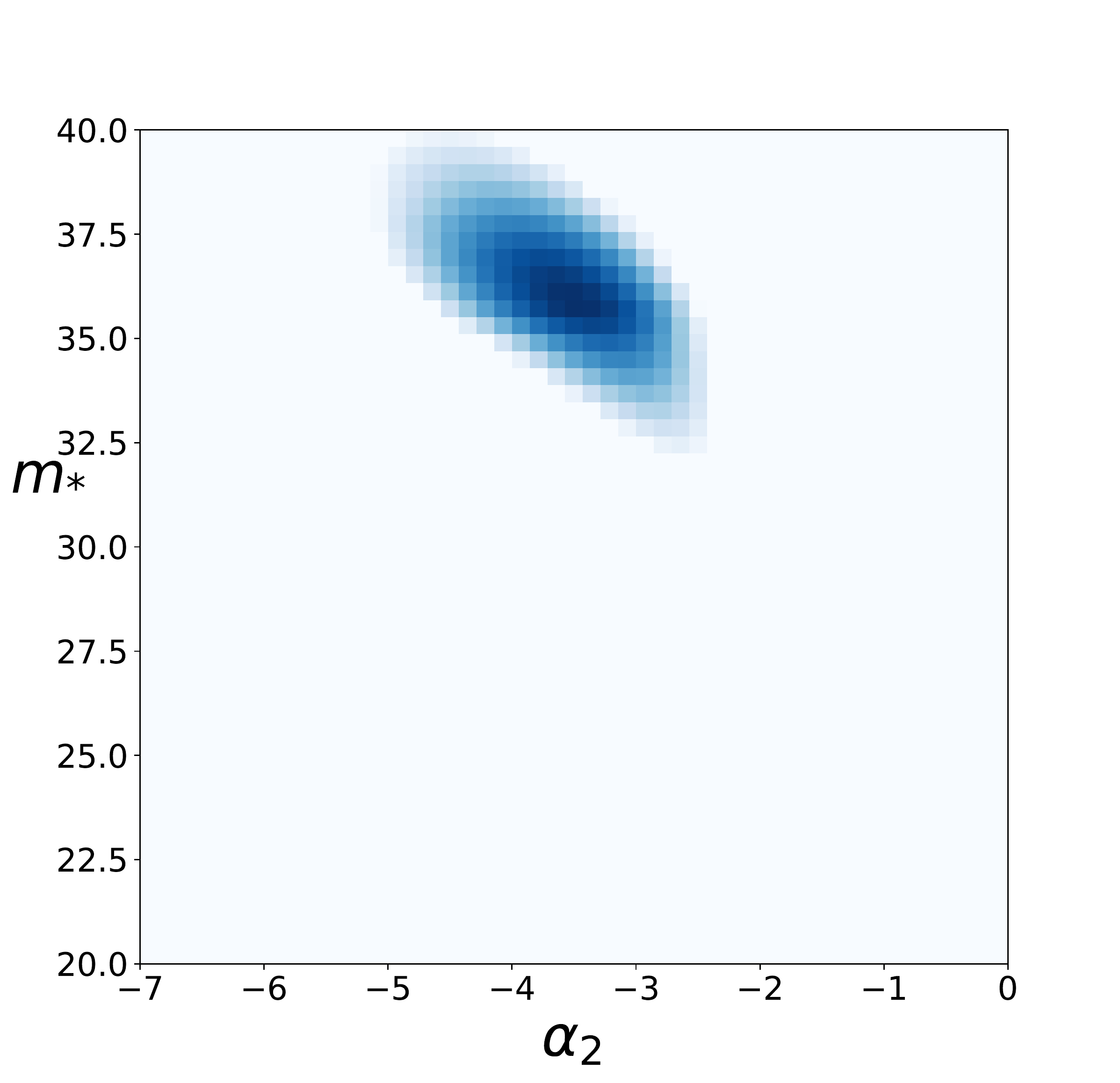}\includegraphics[scale=0.22]{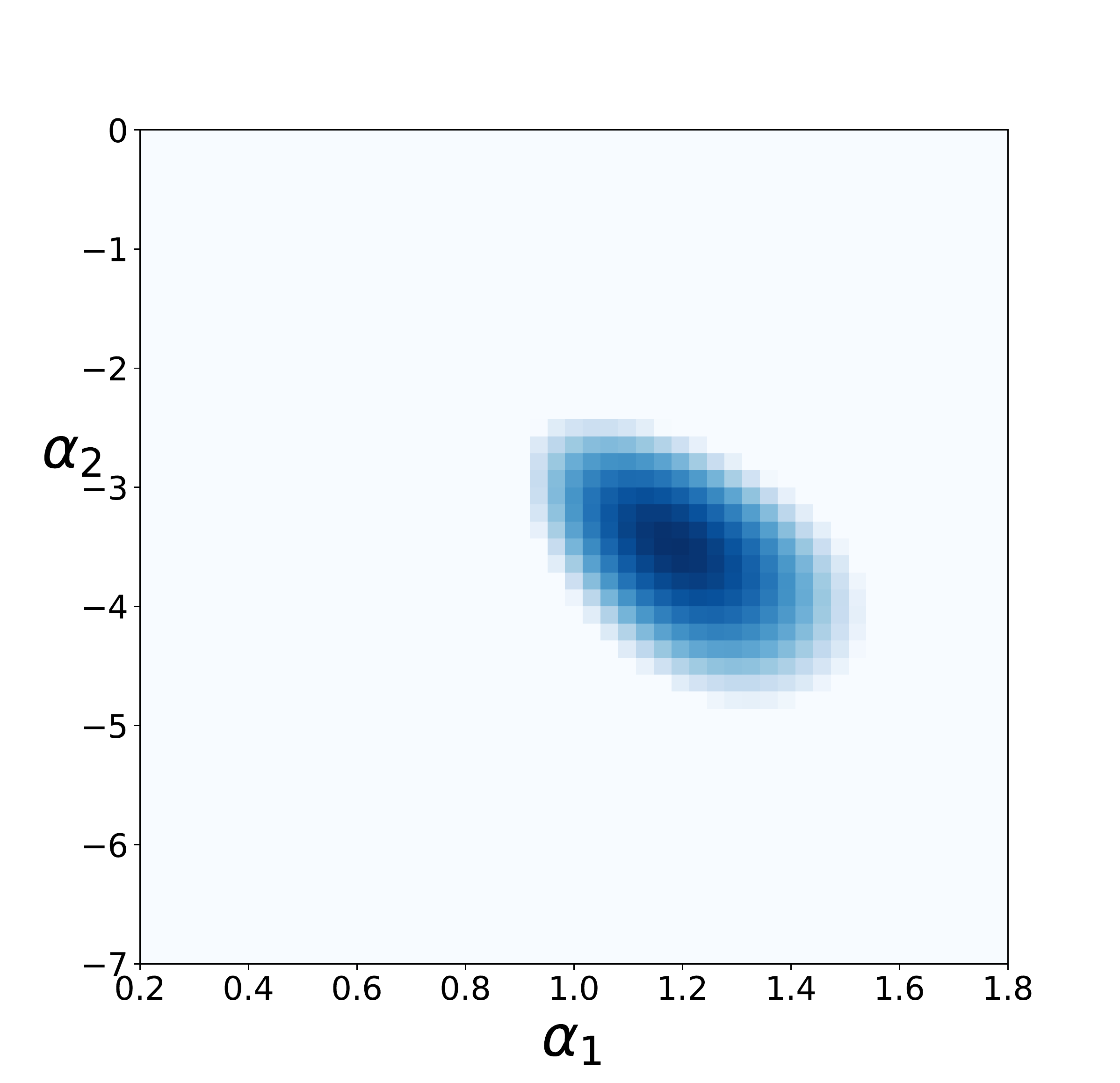}\protect\caption{\label{alpha1alpha2m}Three 2-D slices of the 4-D parameter space
with $f_{PBH}=0.0013$. The darkest spot corresponds to the largest
likelihood $\mathcal{L}_{max}$, and the shaded regions are where
the parameter values give likelihoods $\mathcal{L}$ with $\ln\left(\mathcal{L}_{max}/\mathcal{L}\right)\protect\leq2.$
\textit{Left:} $\alpha_{2}\approx1.2$. \textit{Middle:} $\alpha_{1}\approx-4$.
\textit{Right:} $m_{*}\approx35M_{\odot}.$ The unit for $m_{*}$ is
$M_{\odot}$.}
\end{figure}

There are four parameters in a broken power law (\ref{eq:psi}): $\alpha_{1}$,
$\alpha_{2}$, $m_{*}$ and $f_{PBH}$. Scanning through the relevant
part of the parameter space, we found the largest log-likelihood $\ln\mathcal{L}_{max}$
at
\begin{equation}
\alpha_{1}\approx1.2,\ \alpha_{2}\approx-4,\ m_{*}\approx35M_{\odot},\text{\ }f_{PBH}\approx0.0013.\label{eq:para}
\end{equation}
These results are also partly shown in fig. \ref{alpha1alpha2m}.
The power law mass function from different PBH mechanisms was discussed
in, e.g., refs.\cite{Deng:2016vzb,Deng:2017uwc,DeLuca:2020ioi}, which suggested $f(m)\propto m^{-1/2}$ (for large masses).
These models are obviously disfavored by LIGO-Virgo.

As a comparison, we did the same analysis for the log-normal mass
function 
\begin{equation}
\psi(m)=\frac{f_{PBH}}{m\sqrt{2\pi}\sigma}\exp\left[-\frac{\ln^{2}(m/m_{c})}{2\sigma^{2}}\right].
\end{equation}
The most probable parameters are
\begin{equation}
m_{c}\approx20M_{\odot},\ \sigma\approx0.6,\ f_{PBH}\approx0.0013,
\end{equation}
which are consistent with the results found in refs. \cite{Raidal:2017mfl,Raidal:2018bbj,Wong:2020yig,Hutsi:2020sol}. The difference
between the maximum log-likelihood of broken power law and that of
log-normal is $\ln(\mathcal{L}_{BPL}/\mathcal{L}_{LN})\approx5.$
However, considering that we are evaluating the likelihood ratio with
two different models, and that the broken power law has one more free
parameter, we cannot say with much confidence that the broken power
law is a better model at the moment in explaining the LIGO-Virgo results.

\begin{figure}
\includegraphics[scale=0.3]{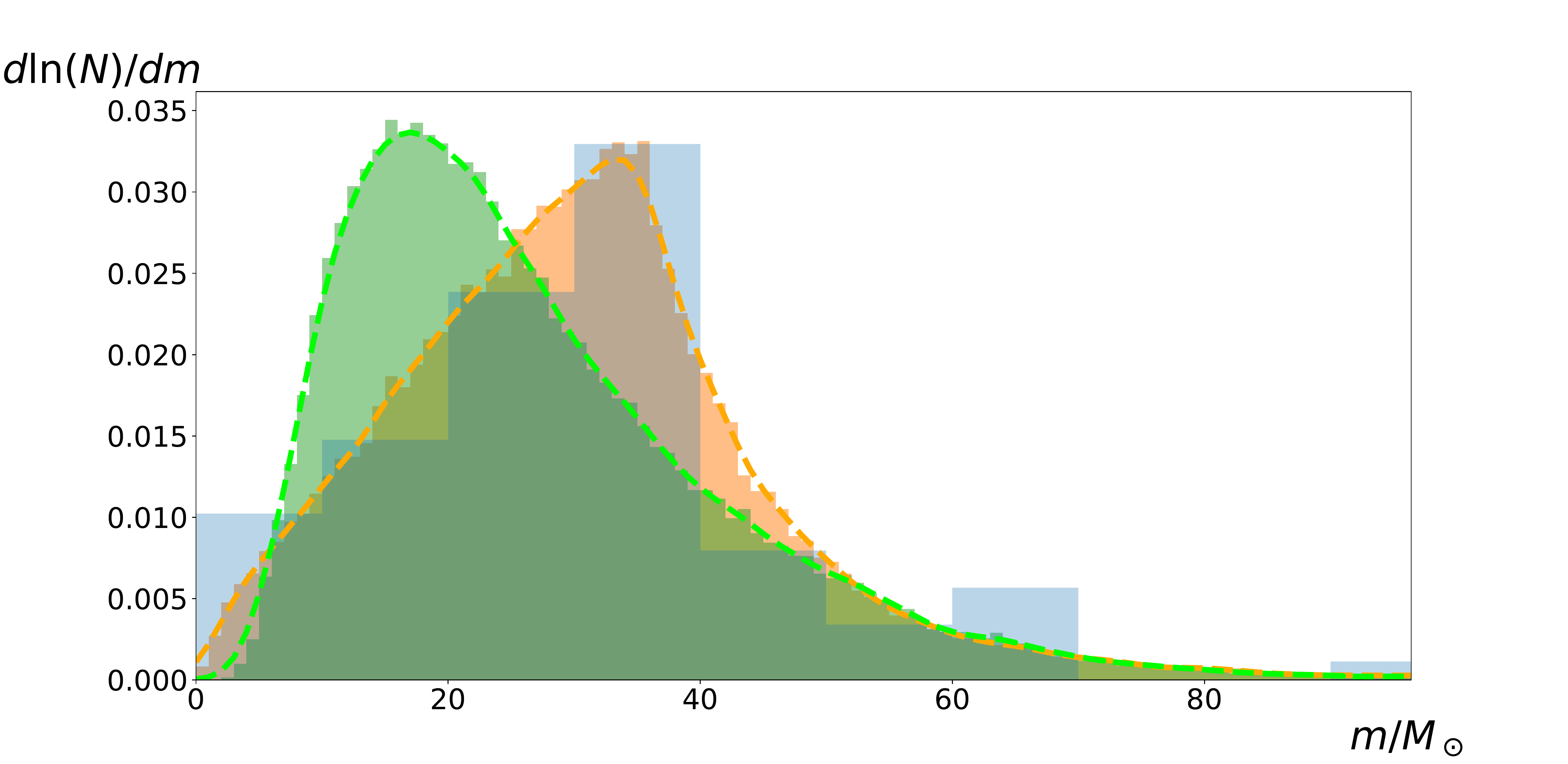}

\caption{\label{fig:LIGO}Mass distributions of black holes from LIGO-Virgo
(blue), the best-fit broken power law (orange) and log-normal (green). }
\end{figure}

Nevertheless, using the Metropolis--Hastings algorithm, we have drawn
50000 random samples from the probability distribution $p_{det}(m_{1},m_{2},z)\text{d}N$
with the best-fit parameters, both for the broken power law and the
log-normal model, and compared the resulting mass distributions with
that from LIGO-Virgo black holes. As we can see from fig. \ref{fig:LIGO},
although it is difficult to compare models under likelihood analysis,
the broken power law mass function appears to be a better fit than
the log-normal.

In the next section we will discuss a physical mechanism where PBHs
could form with such a simple distribution.

\section{\label{PBHs from primordial bubbles} PBHs from primordial bubbles}

In a series of works \cite{Garriga:2015fdk,Deng:2017uwc,Deng:2020mds}, we studied in detail a mechanism of PBH
formation where black holes are formed by vacuum bubbles that nucleate
during inflation. After inflation ends, a bubble will eventually turn
into a black hole in the either subcritical or supercritical regime.
PBHs formed in the two regimes might obey different power laws, and
the transition region could just be the peak at $\sim30M_{\odot}$
seen by LIGO-Virgo. We shall first briefly describe how these black
holes are formed, and then discuss how they are constrained by observations
including LIGO-Virgo. 

\subsection{PBH formation}

Vacuum bubbles could nucleate during inflation via quantum tunneling
if there is a (positive) ``true'' vacuum near the inflationary (quasi-de
Sitter) vacuum in a multidimensional field potential. The bubble interior has an
energy density  $\rho_{b}$ smaller than the inflationary energy density $\rho_{i}$. A typical
bubble expands rapidly, almost at the speed of light. After inflation
ends, inflatons outside the bubble turn into radiation with density
$\rho_{r}\approx\rho_{i}$, and the bubble will run into the surrounding
radiation fluid with a huge Lorentz factor. The bubble continues to
grow, but will eventually come to a halt with respect to the Hubble
flow, because all the forces acting on the bubble wall, including
the interior vacuum pressure, exterior radiation pressure, and the
wall's surface tension, point inwards. A bubble is called ``subcritical''
if it simply collapses into a black hole after reentering the Hubble
horizon. A supercritical bubble, on the other hand, will inflate without
bound due to the repulsive vacuum inside. However, instead of consuming
our universe, the bubble grows into a baby universe, which is connected
to us by a wormhole. For an exterior observer, the wormhole is a spherical
object that will eventually turn into a black hole as the ``throat''
pinches off. Once the relation between the black hole mass in these
two regimes and the bubble's size at the end of inflation is determined, we are
able to obtain the mass spectrum of these black holes from the bubbles' size distribution.

In ref. \cite{Deng:2017uwc}, we considered an ideal scenario where radiation outside
the bubble can be completely reflected by the bubble wall, which implies strong interactions between the bubble field and the standard model particles. In this
setting, the mass function was found to be $f\propto m^{-1/2}$ in the supercritical regime. As we know from
the previous section, such a distribution is disfavored by the LIGO-Virgo
results.

In ref. \cite{Deng:2020mds}, we studied the other extreme possibility, where
the bubble wall is transparent, and radiation can freely flow inside. In this case, the trajectory
of the bubble wall before it ceases to grow with respect to the Hubble
flow can be estimated by assuming an FRW metric dominated by radiation
outside the bubble and matching the spacetimes on two sides of the
wall. Let $r$ be the the bubble's comoving radius. The equation
of motion of the bubble wall can be found to be
\begin{equation}
\ddot{r}+\left(4-3a^{2}\dot{r}^{2}\right)H\dot{r}+\frac{2}{a^{2}r}\left(1-a^{2}\dot{r}^{2}\right)+\left(\frac{\rho_{b}}{\sigma}+6\pi\sigma\right)\frac{\left(1-a^{2}\dot{r}^{2}\right)^{3/2}}{a}=0,\label{chiEOM}
\end{equation}
where the overdot represents the derivative with respect to cosmic time $t$, $H\equiv\dot{a}/a=(2t)^{-1}$
is the Hubble parameter in the exterior FRW universe, and $\sigma$ is the surface tension of the bubble wall. Let $t_{i}$
be the time when inflation ends, then the scale factor is defined
by $a=\left(t/t_{i}\right)^{1/2}$. The ``initial'' conditions of
eq. (\ref{chiEOM}) is $r(t_{i})=r_{i}$ and $\dot{r}(t_{i})=\left(1-\gamma_{i}^{-2}\right)^{1/2}$,
where $r_{i}$ can be smaller or much larger than the Hubble horizon
$H_{i}^{-1}$ at $t_{i}$, and $\gamma_{i}$ (regarded as a free parameter\footnote{See discussion in section \ref{Conclusions}.})
is the Lorentz factor of the bubble wall with respect to the Hubble
flow at $t_{i}$. 

Assuming a huge $\gamma_{i}$, the trajectory of the wall can be approximated
by $a\dot{r}\approx1$. Let $t_{s}$ be
the time when the wall comes to a stop with respect to the Hubble
flow. The bubble's comoving radius at $t_{s}$ can then be estimated
as 
\begin{equation}
r_{s}\equiv r(t_{s})\sim r_{i}+\left(a_{s}-1\right)H_{i}^{-1},\label{chi_s}
\end{equation}
where $a_{s}\equiv\left(t_{s}/t_{i}\right)^{1/2}$. We are particularly
interested in the case where $\gamma_{i}$ is so large that $r_{s}\gg r_{i}$.
In this case, $r_{s}\sim a_{s}H_{i}^{-1}$, and hence the bubble's
physical radius at $t_{s}$ can be estimated as
\begin{equation}
R_{s}\equiv a_{s}r_{s}\sim a_{s}^{2}H_{i}^{-1}\sim t_{s},\label{eq:R_s}
\end{equation}
which means the bubble size is comparable to the Hubble horizon at
$t_{s}.$

\subsubsection{subcritical bubble}

A bubble with sufficiently small $r_{i}$ would continue to expand
after $t_{s}$, until it reaches a maximum physical size $R_{max}$.
It will then shrink and eventually collapse into a black hole. By
eq. (\ref{eq:R_s}), the bubble reenters the horizon at $\sim t_{s}$,
then its evolution will no longer be affected by Hubble expansion
significantly, which means the mass of the bubble itself (excluding
radiation inside) will almost conserve after then. When the bubble radius reaches $R_{max}$, the kinetic energy of the bubble wall vanishes. Furthermore, since the bubble is expected to shrink upon horizon reentry, we have $R_{max} \sim R_s$. Hence, the mass of the resulting black hole from a subcritical bubble is
\begin{equation}
m\sim\frac{4}{3}\pi\rho_{b}R_{s}^{3}+4\pi\sigma R_{s}^{2}-8\pi^{2}\sigma^{2}R_{s}^{3}.\label{eq:sub}
\end{equation}
Here the three terms
can be understood as the volume energy, the surface energy and the surface binding energy, respectively. 

As $r_{i}$ is increased to a critical value, $R_{max}$ can no longer
be reached, because a sufficiently large bubble dominated by its interior vacuum would inflate. When this happens, we enter the supercritical regime.

\subsubsection{supercritical bubble}

Due to the third term on the left hand side in eq. (\ref{chiEOM}),
a bubble with a larger $r_{i}$ experiences a smaller ``friction''
as it grows, and therefore has a larger $t_{s}$. By the time $t_{s}$, the bubble's vacuum density $\rho_{b}$ could become even larger than the exterior radiation density $\rho_{r}(t_s)$ (note that $\rho_{r}(t_{i})=\rho_{i}>\rho_b$), i.e.,
\begin{equation}
\rho_{r}(t_{s})=\rho_{i}a_{s}^{-4}\lesssim\rho_{b}.
\end{equation}
Then by eq. (\ref{eq:R_s}),
\begin{equation}
R_{s}\sim a_{s}^{2}H_{i}^{-1}=\left(\frac{8\pi}{3}\rho_{i}a_{s}^{-4}\right)^{-1/2}\gtrsim\left(\frac{8\pi}{3}\rho_{b}\right)^{-1/2}=H_{b}^{-1}.
\end{equation}
Here $H_{b}^{-1}$ is the de Sitter horizon associated with the bubble
vacuum. $R_{s}\gtrsim H_{b}^{-1}$ then implies that the bubble would
inflate before $t_{s}$ and thus create a wormhole, which later turns
into a black hole. By the estimate in ref. \cite{Deng:2020mds}, the black hole
mass in this supercritical regime is
\begin{equation}
m\sim H_{i}\left(r_{s}-c_{s}a_{s}H_{i}^{-1}\right)^{2},\label{eq:sup}
\end{equation}
where $c_{s}$ is the speed of sound in the radiation fluid, and the
value of $a_{s}$ is determined by solving eq. (\ref{chiEOM}). By the estimate in eq. (\ref{eq:R_s}), we have $m\sim H_{i}r_s^2 \sim a_s^2 H_i^{-1}\approx t_s$, which is the horizon mass at $t_s$.

$\ $

Therefore, by numerically integrating eq. (\ref{chiEOM}) till $\dot{r}=0$, we are
able to find out the black hole masses in the subcritical and the
supercritical regimes. These two regimes are expected to be connected
by a region where the black hole mass transits smoothly from one regime
to the other. In practice, we evaluate both (\ref{eq:sub}) and (\ref{eq:sup}),
and take the smaller one to be the black hole mass for a certain bubble. 

\subsection{Size distribution and mass function}

Now that we have the masses of black holes from bubbles with various $r_{i}$,
we need the distribution of $r_{i}$ in order to obtain the mass
function. Bubbles formed earlier expand to larger sizes,
but they are rarer due to cosmic expansion. By assuming that bubbles
are formed with a vanishing size and that the bubble worldsheet is
the future light cone of the nucleation point, one can find the number
density of bubbles having radius in the interval $(r_{i},r_{i}+\text{d}r_{i})$ at $t$ ($t>t_{i}$) to be \cite{Garriga:2015fdk}
\begin{equation}
\text{d}n(t)\approx\lambda\frac{\text{d}r_{i}}{a(t)^{3}\left(r_{i}+H_{i}^{-1}\right)^{4}}.
\end{equation}
where $\lambda$ is the dimensionless bubble nucleation rate per Hubble
volume per Hubble time. Then by the definition of $f(m)$ (eqs. (\ref{eq:psi def})
and (\ref{eq:f def})), we have
\begin{equation}
f(m)\propto\frac{\lambda m^{2}}{H_{i}^{3/2}\left(r_{i}+H_{i}^{-1}\right)^{4}}\frac{\text{d}r_{i}}{\text{d}m}.\label{f(M)}
\end{equation}
The relation between $r_{i}$ and $m$ can be found by numerically
solving eq. (\ref{chiEOM}) and using eqs. (\ref{eq:sub}) and (\ref{eq:sup}).
Therefore, the mass function is completely determined by the following
five (independent) parameters: the Lorentz factor of the bubble wall
at the end of inflation  $\gamma_{i}$, the bubble nucleation rate $\lambda$, the
inflationary density $\rho_{i}$, the vacuum density of the bubble
interior $\rho_{b}$ , and the bubble wall tension $\sigma$. In the
following we shall use
\begin{equation}
\eta_{i}\sim\rho_{i}^{1/4},\ \eta_{b}\sim\rho_{b}^{1/4},\ \eta_{\sigma}\sim\sigma^{1/3},
\end{equation}
to characterize $\rho_{i}$, $\rho_{b}$ and $\sigma$. They represent
the energy scales of inflation, bubble interior and bubble wall, respectively. 

\begin{figure}
\includegraphics[scale=0.5]{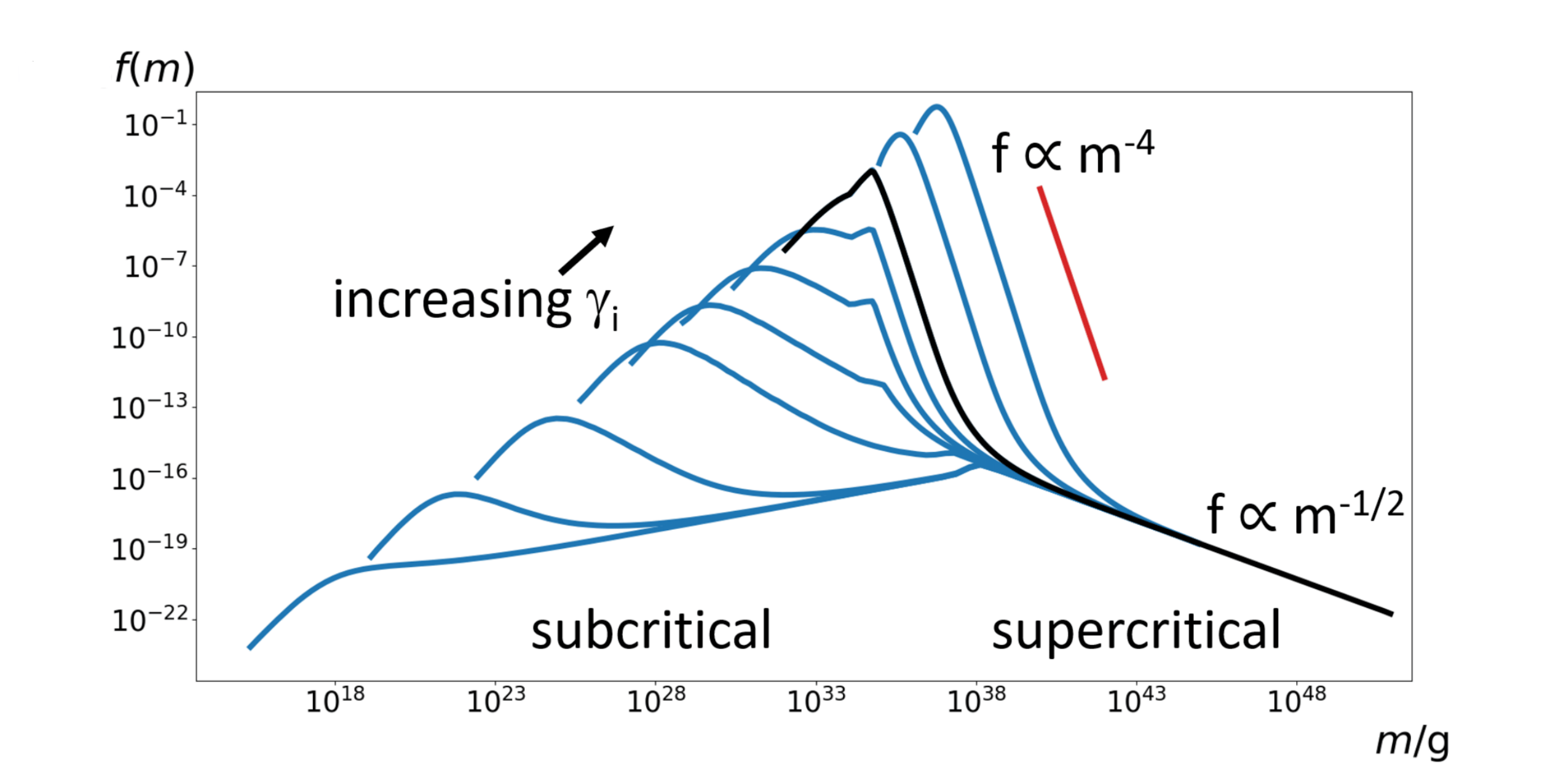}

\caption{\label{fig:limit1+2}Some examples of the mass function $f(m)$ given
by (\ref{f(M)}). We have fixed parameters except for the Lorentz
factor $\gamma_{i}$, with $\lambda\approx10^{-21},$ $\eta_{i}\approx 10^{-16}M_{Pl}$, $\eta_{b}\approx 6\times 10^{-21}M_{Pl},$
 $\eta_{\sigma}\approx 4\times 10^{-15}M_{Pl}$, and $\gamma_{i}$ varying from $10^{4}$ to $10^{30}$. For small $\gamma_{i}$ the maximum of $f(m)$
appears at the transition region. As $\gamma_{i}$ increases, another
peak develops in the subcritical regime. For a sufficiently large
$\gamma_{i}$, supercritical black holes near the transition region approximately follow
a power law $f\propto m^{-4}$ (the red straight line is $\propto m^{-4}$). For large masses, all spectra approach
$f\propto m^{-1/2}$. The mass function in black (the third one from the right) agrees well with the broken power law suggested by LIGO in the relevant mass range, as shown in fig. \ref{fig:f_PBH}.}
\end{figure}

Depending on the parameter values, the mass function (\ref{f(M)})
can have very different shapes and can be wide or relatively narrow.
Several examples of $f(m)$ are shown in fig. \ref{fig:limit1+2},
where we have fixed all other parameters except for the Lorentz factor
$\gamma_{i}$.\footnote{Note that other parameters are not less essential than $\gamma_i$. Varying them will give different curves of $f(m)$ (shifting positions on the $f\mbox{--}m$ plane) that look similar to those in fig. \ref{fig:limit1+2}.} An intriguing feature relevant to our discussion
is that when $\gamma_{i}$ is sufficiently large, black holes in the supercritical regime
near the transition approximately follow a power law $f(m)\propto m^{-4}$ (by a numerical fit to the curves)\footnote{A semi-analytic discussion of this relation can be found in the appendix. For supercritical bubbles with a huge $\gamma_i$, the mass function near the transition satisfies $f\propto m^{-4.25}$.},
which is what LIGO-Virgo implies for PBHs with
$m>m_{*}\sim35M_{\odot}.$ Moreover, there is a peak in the transition
region for some values of $\gamma_{i}$, and the mass function can
be approximated by different power laws on two sides of the peak.
Now the question is whether suitable bubble parameters can be found
such that subcritical black holes obeys $f(m)\propto m^{1.2}$.

\subsection{\label{Observational constraints}Observational constraints}

\begin{figure}
\includegraphics[scale=0.3]{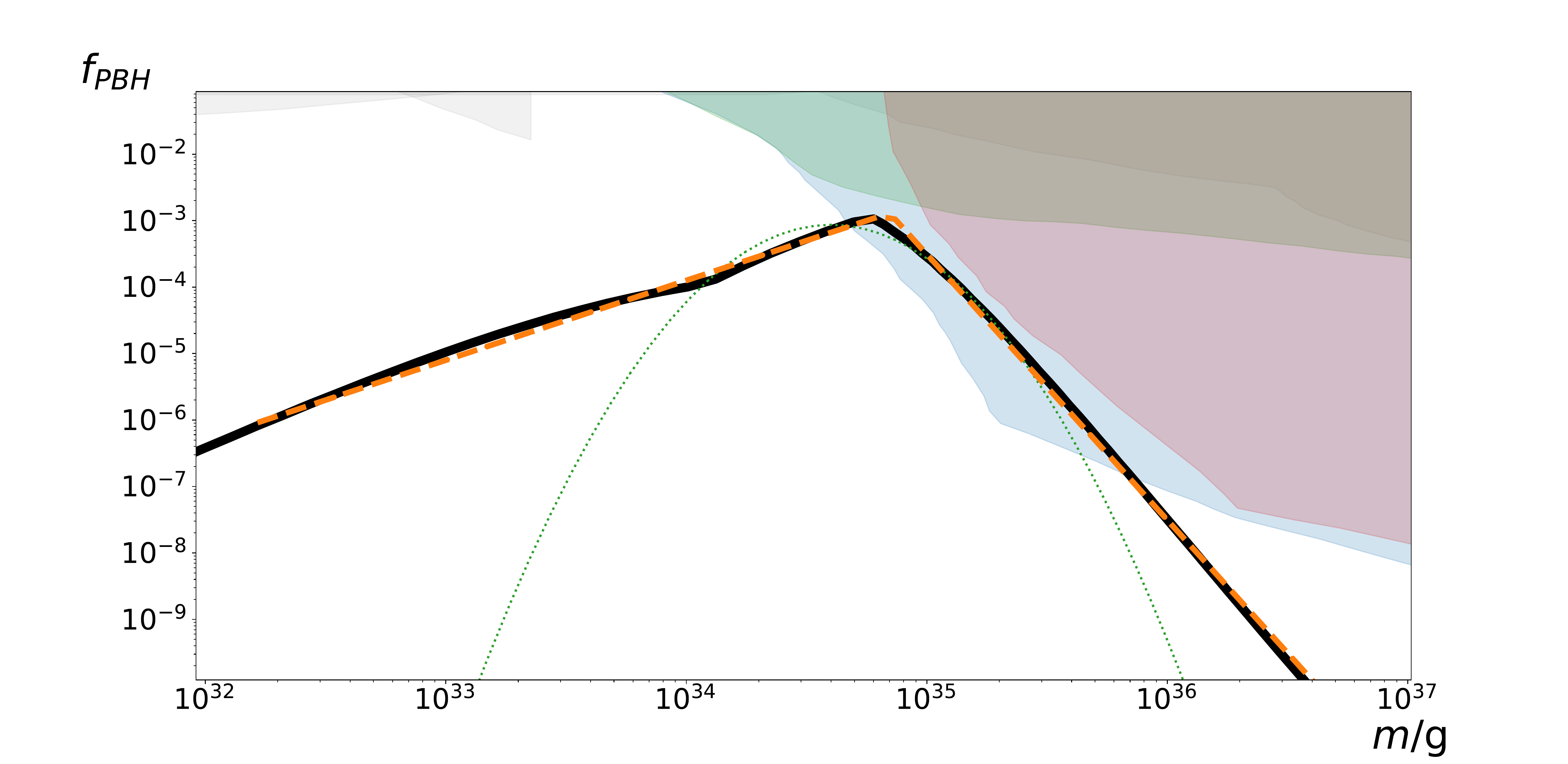}

\caption{\label{fig:f_PBH}Observational constraints on our mechanism. The light grey, light blue and light purple regions are
excluded for PBHs with a monochromatic spectrum \cite{Alcock:2000ph,Tisserand:2006zx,Oguri:2017ock,Authors:2019qbw,Inoue:2017csr,Serpico:2020ehh} (adapted from fig. 10 in ref. \cite{Carr:2020gox}). The light green region is a modification of the light blue, with the effect of PBH accretion taken into account \cite{DeLuca:2020fpg}. The black solid curve is a possible distribution
from our PBH mechanism. 
The orange dashed lines are the best-fit broken power law,
with $f\propto m^{1.2}$ for $m<35M_{\odot}$ and $f\propto m^{-4}$
for $m>35M_{\odot}$. The green dotted curve is the best-fit log-normal mass function.}
\end{figure}

The answer is yes. In fact, there are many sets of bubble parameters
that can give the desired mass function. In fig. \ref{fig:f_PBH}
we show an example of $f(m)$ with $\gamma_{i}\approx10^{22},$
$\lambda\approx10^{-21},$ $\eta_{i}=\mathcal{O}(1)\ \text{TeV}$, $\eta_{b}=\mathcal{O}(0.1)\ \text{GeV}$
and $\eta_{\sigma}=\mathcal{O}(10)\ \text{TeV}$ (black, solid curve). It agrees well with
the broken power law mass function with the parameters in eq. (\ref{eq:para})
(orange, dashed lines). As a comparison, we also show the best-fit
log-normal function (green, dotted curve). The light grey and colored
areas are regions excluded by different astrophysical observations  (see ref. \cite{Carr:2020gox} for a recent
review).
Strictly speaking, these upper bounds are valid only for PBHs with
a very narrow spectrum, and so are not supposed to be used to compare
with an extended mass function $f(m)$. (The method of applying these
bounds on a broad spectrum was discussed in ref. \cite{Carr:2017jsz}.) Loosely
speaking, however, as long as $\psi(m)=f(m)/m$ does not have a plateau
over a relatively big range, we can still constrain a model by placing
$f(m)$ near the bounds while avoiding hitting them. 

The most stringent bound for us in fig. \ref{fig:f_PBH},
which covers the mass range around $30\mbox{--}300M_{\odot}$, comes
from Planck. Ref. \cite{Serpico:2020ehh} studied how disk or spherical accretion of
a halo around PBHs could affect CMB, which strongly
constrains the population of PBHs in the range around $1\mbox{--}10^{4}M_{\odot}$.
The broken power law function implied by LIGO-Virgo seems in
tension with the model of disk accretion (light blue), but is free
from the bound imposed by spherical accretion (light purple). However, we note that the effect of PBH accretion was not taken into account in our discussion. It is pointed
out in refs. \cite{DeLuca:2020fpg,DeLuca:2020sae,DeLuca:2020qqa} that efficient PBH accretion before the reionization
epoch tends to relax the tension with these bounds.\footnote{Although the process of PBH accretion is model-dependent, and is even more uncertain for binary systems, it is generally agreed that PBHs with masses smaller than $\mathcal{O}(10)M_\odot$ are not significantly affected by accretion, and so have tiny spins, while larger PBHs generally have larger spins. This is compatible with the LIGO-Virgo results that the correlation between mass and the mean
effective spin is negative, whereas the correlation between mass and the spin dispersion is positive \cite{Safarzadeh:2020mlb}. The best-fit of a one-parameter accretion model with the GWTC-1 and GWTC-2 catalogs can be found in ref. \cite{Wong:2020yig}.} For instance, if one adopts the one-parameter (denoted by $z_{\text{cut-off}}$) accretion model in ref. \cite{DeLuca:2020fpg} (here with $z_{\text{cut-off}}=10$), the light blue constraint in fig. \ref{fig:f_PBH} is relaxed to the light green region, which does not intersect with our mass function.

Besides being able to account for the LIGO-Virgo black holes, another interesting feature of our mass function is that it may provide
an explanation to supermassive black holes (SMBHs) at the center
of most galaxies \cite{LyndenBell:1969yx,Kormendy:1995er}. These black holes have masses from $\sim 10^{6}\text{--}10^{10}M_{\odot}$,
which cannot be explained by standard accretion models \cite{Haiman:2004ve}. Moreover,
observations of quasars indicate that many of them were already in
place at high redshifts (see ref. \cite{Kormendy:2013dxa} for a review). Recently,
a black hole as large as $\sim10^{9}M_{\odot}$ was discovered at $z\approx7.6$,
which greatly challenges the conventional formation theory of astrophysical
black holes \cite{wang2021luminous}. One is then led to speculate that SMBHs were seeded by
PBHs, which could have large masses by birth \cite{Rubin:2001yw,Bean:2002kx,Duechting:2004dk,Clesse:2015wea,Carr:2018rid}. At the
present time ($t_{0}$), the number density of PBHs of mass $\sim m$
is approximately given by
\begin{equation}
n(m)\sim\rho_{CDM}(t_{0})\frac{f(m)}{m}\approx4\times10^{11}\left(\frac{M_{\odot}}{m}\right)f(m)\ \text{Mpc}^{-3}.
\end{equation}
It was shown in refs. \cite{Serpico:2020ehh} that PBHs at $z\sim6$ with $m \sim 10^{2}\text{--}10^{4}M_{\odot}$
can attain sufficient accretion and grow to SMBHs even if $f(m)$
is as small as $10^{-9}$. 
We can see from fig. \ref{fig:f_PBH} that the best-fit broken power law mass function gives $f(10^{3}M_{\odot})\gtrsim10^{-9}$,
which gives $n(10^{3}M_{\odot})\sim0.4\ \text{Mpc}^{-3}$. This is
larger than the present density of galaxies $\sim0.1\ \text{Mpc}^{-3}$. Therefore, SMBHs could indeed have been seeded by these PBHs.

\section{\label{Conclusions} Conclusions and discussion}

In this work we have used maximum likelihood estimation to analyze
the GWTC-1 and GWTC-2 datasets released by the LIGO-Virgo Collaboration,
assuming that all of the black holes in the 44 confident merger events
are PBHs with a mass spectrum $f(m)$ satisfying a
broken power law. We found that the best-fit $f(m)$ has a peak at $m_{*}\approx35M_{\odot}$, consistent
with the observations, and that $f\propto m^{1.2}$ for $m<m_{*}$ and $f\propto m^{-4}$
for $m>m_{*}$. These black holes can constitute about $0.1\%$ of
the dark matter. 

Such a unique mass function can be realized in a mechanism of PBH
formation we proposed in recent years, where the black holes are formed by vacuum bubbles that could nucleate during inflation. In general, the inflaton field rolls in a multidimensional potential containing a number of minima. Bubble nucleation can occur during inflation if a minimum has energy scale lower than the inflationary scale. After inflation ends, depending on their size, these
bubbles will turn into black holes either by simple collapse (subcritical),
or by creating a wormhole (supercritical). The resulting mass function
of the black holes may obey different power laws in the two regimes,
which are connect by a transition region that could possibly be at $m_{*}$.
Surprisingly, if the bubble walls have a sufficiently large Lorentz factor
at the end of inflation, which is typically the case, the mass function
in the supercritical regime near $m_{*}$ is indeed approximately given by $f\propto m^{-4}$. With properly chosen parameters (including the Lorentz factor, the energy scales of bubble vacuum, bubble wall and inflation), we can have $f\propto m^{1.2}$ for $m<m_{*}$. 

With this mass function, we can also have a sufficient number of seeds $(m\sim 10^3 M_\odot)$ that could grow into SMBHs observed at the galactic centers, which is difficult to explain with conventional accretion models. In addition, as we can see in fig. \ref{fig:limit1+2}, it is also possible to estimate the distribution of stupendously large black holes \cite{Carr:2020erq} from the large mass end, where $f\propto m^{-1/2}$. This however will not be discussed further in the present work.

Another noticeable feature of our mass function is that it suggests a relatively low inflationary scale. During the slow roll, the ultimate Lorentz factor of the bubble wall for a comoving observer outside the bubble is $\gamma \sim \eta_i^2 M_{Pl} / \eta^{(i)3}_\sigma$, where $\eta^{(i)}_\sigma$ is the energy scale of the  wall tension $\sigma$ during inflation \cite{Berezin:1987bc,Deng:2020mds}. We assumed $\gamma$ to be the Lorentz factor of the wall as it runs into the ambient radiation after inflation, i.e., $\gamma_i=\gamma$. However, at the end of inflation, the inflaton quickly rolls down into our vacuum, which could cause a drastic change in $\sigma$ since the shape of the barrier in the field potential could change significantly. Therefore, we regarded $\gamma_i$ and $\sigma$ as two free parameters. With the parameter values that lead to the black solid mass function in fig. \ref{fig:f_PBH}, where the inflationary scale is $\eta_i=\mathcal{O}(1)\ \text{TeV}$, we have $\eta_\sigma/\eta^{(i)}_\sigma=\mathcal{O}(100\text{--}1000)$. If we require a larger $\eta_i$, the ratio $\eta_\sigma/\eta^{(i)}_\sigma$ is even larger, which seems less likely. Of course, this should be determined by the configuration of the multidimensional potential, which is beyond the scope of the present work. 

An inflationary model at
the TeV scale was constructed in ref. \cite{Copeland:2001qw}. In this model of hybrid inflation, the inflaton is directly
coupled to the Higgs field such that the symmetry is restored even
at a temperatures lower than the electroweak scale.\footnote{New fields need to be introduced in order to produce bubbles in our model.} As the Higgs potential
becomes unstable later, the fields roll down in random directions, leading
to non-trivial Higgs configurations. In the presence of CP-violation,
this might produce a baryon asymmetry, which is referred to as cold
electroweak baryogenesis \cite{GarciaBellido:1999sv,Krauss:1999ng,Smit:2002sn,Konstandin:2011ds}, and has been studied extensively
with simulations \cite{Tranberg:2006ip,Tranberg:2006dg,Mou:2017atl,Mou:2018xto}. Such a process might also be a source of
primordial magnetic fields \cite{DiazGil:2008tf,Mou:2017zwe} and stochastic gravitational waves \cite{GarciaBellido:2007af}.

In hybrid inflation, the inflationary scale could be so small that the expansion rate is much smaller than the typical particle decay rate, hence thermalization occurs almost instantaneously compared to Hubble time \cite{GarciaBellido:1999sv}. This is compatible with the assumption in our model where the bubble runs into radiation immediately after inflation ends. If this is not the case, a different matter content during reheating would change the scale factor in the equation of motion of the bubble wall (eq. (\ref{chiEOM})). To see how this would affect the resulting mass function, we solved the equation assuming the universe is dust-dominated for the first few Hubble times. It turned out the desired mass function (in fig. \ref{fig:f_PBH}) can be obtained by, e.g., increasing $\gamma_{i}$ and $\eta_{\sigma}$ (by a factor $\mathcal{O}(1\mbox{--}10)$) without changing other parameters.

With many more merger events to be detected by LIGO-Virgo in the near future, the mass distribution of black holes can be determined with increasing certainty. Excited as we are, that our mechanism is the only factory of LIGO-Virgo black holes can be ruled out if the mass function turns out to have a very different shape, such as one with a slope (in the log-log plot) much smaller than -4 for large black holes.

\section*{Appendix: Mass distribution of supercritical black holes}

In this appendix, we will find a semi-analytic solution to the bubble
wall's equation of motion for relatively large
bubbles, and thus determine the mass function of supercritical black
holes, which was found numerically in section \ref{PBHs from primordial bubbles}, and is shown in
fig. \ref{fig:limit1+2}. 

The bubble wall's equation of motion after inflation is (eq. (\ref{chiEOM}))
\begin{equation}
\ddot{r}+\left(4-3a^{2}\dot{r}^{2}\right)H\dot{r}+\frac{2}{a^{2}r}\left(1-a^{2}\dot{r}^{2}\right)+K\frac{\left(1-a^{2}\dot{r}^{2}\right)^{3/2}}{a}=0,\label{chiEOM-1}
\end{equation}
where we have defined $K\equiv \rho_{b}/\sigma+6\pi\sigma$. This equation can be rewritten as
\begin{equation}
\dot{u}+\frac{3}{2t}u+\frac{2\gamma}{ar}+K=0,\label{eq:u}
\end{equation}
where $u(t)\equiv\sqrt{\gamma^{2}-1}$ and $\gamma\equiv\left(1-a^{2}\dot{r}^{2}\right)^{-1/2}$.
As discussed in section \ref{PBHs from primordial bubbles}, if we assume a huge $\gamma_{i}$ (or
$u(t_{i})$), the trajectory of the wall can be approximated by $a\dot{r}\approx1$,
which gives
\begin{equation}
r(t)\approx r_{i}+2\sqrt{t_{i}}(\sqrt{t}-\sqrt{t_{i}}).\label{chi}
\end{equation}
Then when the wall comes to a stop with respect to the Hubble flow,
we have
\begin{equation}
r_{s}\sim r_{i}+a_{s}H_{i}^{-1},\label{chi_s-1}
\end{equation}
where we have used $r_{i}\gg H_{i}^{-1}$, which is typically the
case for supercritical bubbles in our discussion. In what follows we shall discuss two limits: (a) $r_{i}\gg a_{s}H_{i}^{-1}$,
and (b) $r_{i}\ll a_{s}H_{i}^{-1}$.

\subsubsection*{(a) $r_{i}\gg a_{s}H_{i}^{-1}$}

In this limit, $r_{s}\sim r_{i}$. Then by eq. (\ref{eq:sup}), the resulting
black hole mass is
\begin{equation}
m\propto r_{i}^{2}.
\end{equation}
Then it can easily be shown that the mass function (eq. (\ref{f(M)}))
becomes
\begin{equation}
f(m)\propto\frac{m^{2}}{r_{i}^{4}}\frac{\text{d}r_{i}}{\text{d}m}\propto m^{-1/2}.\label{f(M)-1}
\end{equation}
This explains the behavior of $f(m)$ on the end of large masses,
as shown in fig. \ref{fig:limit1+2}.

\subsubsection*{(b) $r_{i}\ll a_{s}H_{i}^{-1}$}

In this limit, the bubbles are assumed to still be in the supercritical
regime, but their initial radius $r_{i}$ is negligible compared to
$r_{s}\sim a_{s}H_{i}^{-1}$. By eq. (\ref{chi}), we can approximate
eq. (\ref{eq:u}) by
\begin{equation}
\dot{u}+\frac{3}{2t}u+\frac{2u}{r_{i}\sqrt{t}/\sqrt{t_{i}}+2t}+K=0.\label{eq:u-1}
\end{equation}
Now we can roughly divide the time interval between $t_{i}$ and $t_{s}$
by two stages: before and after $t\approx r_{i}^{2}/4t_{i}\equiv t_{m}$,
which determines whether $r_{i}\sqrt{t}/\sqrt{t_{i}}$ or $2t$ dominates
the denominator in the third term of the above equation. 

Before $t_{m}$, we have
\begin{equation}
\dot{u}+\frac{3}{2t}u+\frac{2\sqrt{t_{i}}}{r_{i}\sqrt{t}}u+K=0,
\end{equation}
which has an analytic solution
\begin{align}
u(t) & =\frac{e^{-4\sqrt{t_{i}t}/r_{i}}}{t^{3/2}}\left[\frac{\gamma_{i}t_{i}^{3/2}}{e^{-4t_{i}/r_{i}}}+K\frac{\Gamma(5,-4t_{i}/r_{i})-\Gamma(5,-4\sqrt{t_{i}t}/r_{i})}{512(\sqrt{t_{i}}/r_{i})^{5}}\right].\ \ (t<t_m)
\end{align}
Here we have taken into account the initial condition $u(t_{i})=\gamma_{i}$.
Then at $t_{m}$, we have
\begin{equation}
u(t_{m})=\frac{e^{-2}}{t_{m}^{3/2}}\left[\frac{\gamma_{i}t_{i}^{3/2}}{e^{-4t_{i}/r_{i}}}+K\frac{\Gamma(5,-4t_{i}/r_{i})-\Gamma(5,-2)}{512(\sqrt{t_{i}}/r_{i})^{5}}\right]\label{eq:u(t_m)}
\end{equation}

After $t_{m}$, eq. (\ref{eq:u-1}) becomes
\begin{equation}
\dot{u}+\frac{5}{2t}u+K=0,
\end{equation}
with the solution
\begin{equation}
u(t)=u(t_{m})\left(\frac{t_{m}}{t}\right)^{5/2}+\frac{2}{7}Kt\left[\left(\frac{t_{m}}{t}\right)^{7/2}-1\right].\ \ (t>t_m)\label{eq:u(t_m)-1}
\end{equation}
At $t_{s}$, we have $u(t_{s})=0$. After some algebra, substituting
$u(t_{m})$ by eq. (\ref{eq:u(t_m)}) in eq. (\ref{eq:u(t_m)-1})
leads to
\begin{equation}
t_{s}\approx 0.5\left(\frac{\gamma_{i}}{K}\right)^{2/7}t_{i}^{1/7}r_{i}^{4/7}.
\end{equation}
In the computations we have assumed
the relation $\gamma_{i}>10^{-3}Kr_{i}^{5}/t_{i}^{4}$,
which is true in the cases we are interested in. This estimate for
$t_{s}$ has also been verified numerically.

Now by eq. (\ref{eq:sup}), we have
\begin{equation}
m\sim t_{s}\propto r_{i}^{4/7}.
\end{equation}
Then the mass function becomes
\begin{equation}
f(m)\propto\frac{m^{2}}{r_{i}^{4}}\frac{\text{d}r_{i}}{\text{d}m}\propto m^{-4.25}.\label{f(M)-1-1}
\end{equation}
This explains the behavior of $f(m)$ in the supercritical regime
near the transition, as shown in fig. \ref{fig:limit1+2}. Note that
this relation is but an approximation. For a smaller $\gamma_{i}$,
it was numerically found that the power for $m$ gets slightly larger,
giving $f\propto m^{-4},$ as shown in fig. \ref{fig:f_PBH}.

\section*{Acknowledgments}
I would like to thank Tanmay Vachaspati, Ville Vaskonen, Alex Vilenkin and the anonymous referee for stimulating discussion and comments. This work
is supported by the U.S. Department of Energy, Office of High Energy Physics, under Award
No. de-sc0019470 at Arizona State University.

\bibliography{LIGO}
\end{document}